# Flying spin qubits

Vanita Srinivasa[a] and Jeremy Levy[b]*
Department of Physics and Astronomy, University of Pittsburgh, Pittsburgh,
Pennsylvania 15260, USA

C. Stephen Hellberg[c]
Code 6390, Center for Computational Materials Science, Naval Research Laboratory,
Washington, DC 20375, USA

[a]e-mail: vas9@pitt.edu
[b]e-mail: jlevy@pitt.edu
[c]e-mail: hellberg@nrl.navy.mil
*author to whom correspondence should be addressed

## Abstract

We present a method for encoding and transporting qubits within a dimerized Heisenberg spin-1/2 chain. Logical qubits are localized at the domain walls that separate the two possible dimerized states. The domain walls can be moved to produce "flying spin qubits". The topological nature of these states makes them stable against a wide class of perturbations. Pairs of domain walls can be used to generate Einstein-Podolsky-Rosen pairs of entangled qubits. We discuss speed limitations within an exactly solvable three-spin model and describe a possible physical realization using quantum dot arrays.

The need for "flying qubits,"[1] i.e., a mechanism for rapidly transporting quantum information, has long been recognized as a weakness of spin-based quantum computing architectures[2-8]. Several different methods have been proposed to implement long-range transport of quantum information. One method involves coupling spin qubits to an external "quantum bus," e.g., an optical cavity mode[5]. Such an approach typically introduces an entirely new set of constraints, and engineering strong optical couplings can be challenging in practice. Another approach attempts quantum information transfer along exchange-coupled spin chains[9-12]. While theoretically possible, the fidelity of transmission depends critically on the values of the coupling strengths between the spins in the chain, making this approach susceptible to errors. Finally, the quantum teleportation protocol[13] of Bennett et al. uses the generation and transport of Einstein-Podolsky-Rosen (EPR) pairs to teleport qubits to their needed location.

Here we present a qualitatively different approach to the construction of "flying qubits". The method relies on the design of a spin-based "designer quantum field", constructed from a one-dimensional dimerized Heisenberg spin-1/2 chain. Logical qubits are localized at the domain walls that separate the two possible dimerized states. Unlike previous encoding schemes, logical qubits are not associated with a definite number of spins; rather, these "defect" states exist even when the number of spins approaches infinity. Their topological nature[14] makes them stable against a wide class of perturbations. By moving the domain walls, it is possible to produce "flying spin qubits". Below, we explore the properties of this class of systems both analytically for small chains (3 spins) and numerically for larger chains (up to 30 spins). Numerical studies for the larger chains are carried out using the Lanczos method of diagonalization[15].

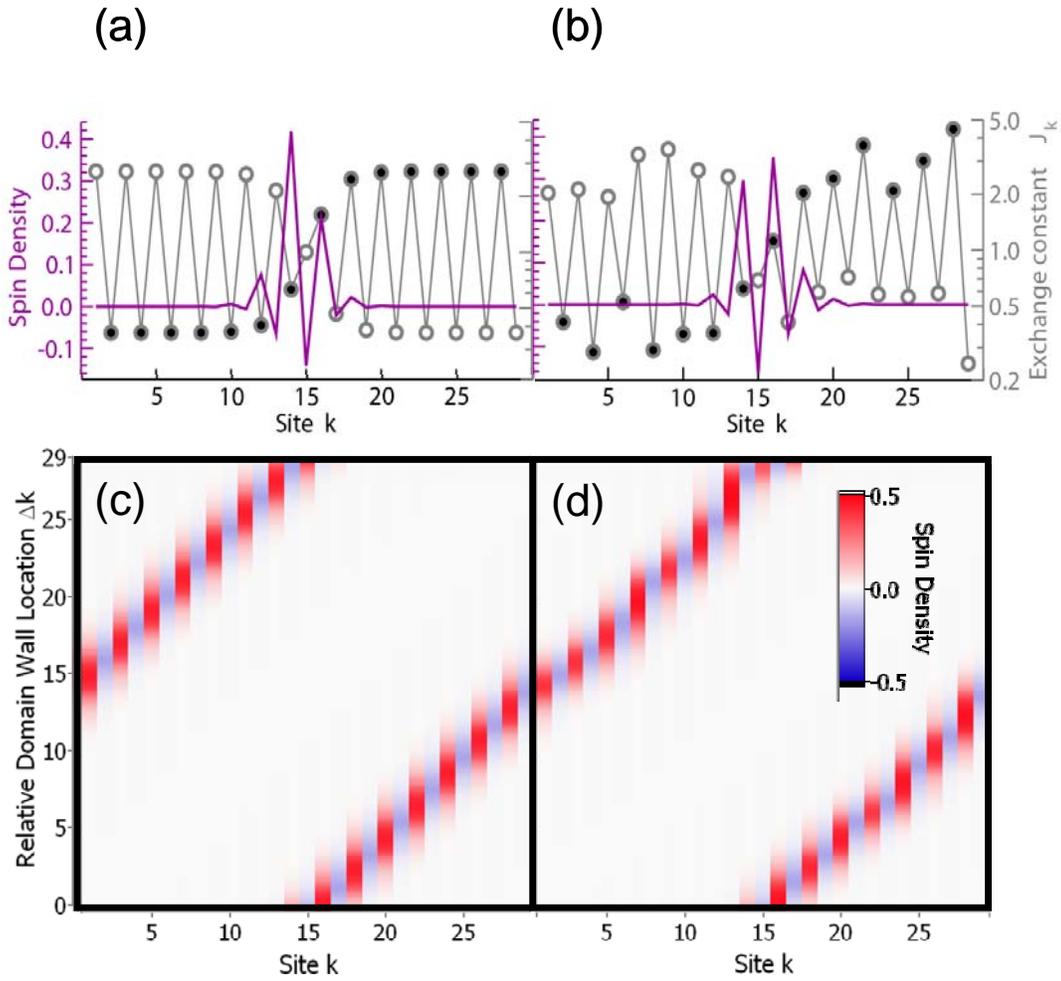

**Figure 1: Generation of a flying spin qubit.** (a) Exchange profile and spin density corresponding to the spin-up ground state for a $n_c = 29$ spin ring containing a single soliton-like state. The filled circles represent even spin sites, and the empty circles represent odd spin sites. (b) Exchange profile and spin density for the spin-up ground state obtained when 50% multiplicative disorder is introduced into the exchange interaction strengths of the system in (a). (c) Adiabatic displacement of the qubit state shown in (a), achieved by shifting the domain wall center position by an amount $\Delta k$ relative to its initial position $k_0 = 15$. The ground state is separated from the first excited state by a gap $\sim J$ for all domain wall locations. (d) Adiabatic displacement of the qubit state for the disordered case, shown in (b).

A one-dimensional system of $n_c$ spin-½ objects interacting via nearest-neighbor Heisenberg exchange is described by the following effective Hamiltonian:

$$H = \sum_{k=1}^{n_c} J_k (\mathbf{S}_k \cdot \mathbf{S}_{k+1}) \equiv \sum J_k (\mathbf{S}_k \cdot \mathbf{S}_{k+1}) \qquad (1)$$

where $\mathbf{S}_k = (X_k, Y_k, Z_k)$ are Pauli operators for the $k^{\text{th}}$ spin, $\{J_k > 0\}$ quantify the strength of nearest-neighbor exchange interactions, and periodic boundary conditions apply ($k \pm n_c \equiv k$). The Hamiltonian in Eq. (1) can be mapped onto a variety of quantum field theories[16] via the Jordan-Wigner spin-particle mapping[17]. We consider a specific parameterization of this Hamiltonian $J_k = J_0 \exp[(-1)^k \alpha(k)]$, where $\alpha$ is a staggered order parameter describing dimerization in the chain. Spatially localized qubits are produced at the zero crossings of $\alpha$, and represent particle-like excitations of a "designer quantum field". The spectrum resembles that of a semiconductor with a soliton-like "defect" state which is closely related to those in conducting polymers[14]. Continuum field-theoretic descriptions[18] are also applicable to these dimerized spin systems, provided the defect extends over many spin lattice sites.

We illustrate the localization of the qubits by considering a closed chain with $n_c = 29$ and $J_0 = 1$ (Figure 1a). A single domain wall is centered at $k_0$ with width $w$ for

$$\alpha(k - k_0) = a_0 \sum_{r=-\infty}^{\infty} (-1)^r \tanh\left((k - (k_0 + r\, n_c))/w\right),$$ where $a_0 = 1$, $k_0 = 15$, and $w = 2$.

The spin density of the spin-up ground state is determined by numerical Lanczos diagonalization and is superimposed on the exchange profile, showing that localization coincides with the zero-crossing of $\alpha$. This localization is retained when 50% multiplicative disorder is introduced into the exchange interaction strengths (Figure 1b), illustrating the topological stability of the qubit with respect to perturbations of the exchange strengths.

By moving the domain wall, it is possible to transport logical spin qubits in a way that preserves the quantum information. Movement of the qubit state around the entire ring, achieved by letting $k_0 \to k_0 + \Delta k$ and varying $\Delta k$, is shown in Figure 1(c). To visualize the qubit, its spin density is plotted as a function of lattice site and domain wall displacement $\Delta k$. Note that two revolutions are required to achieve full periodicity for an $n_c = $ odd spin ring. Variations in the energy gap $E_1 - E_0$ between the ground state $E_0$, which represents the qubit, and the first excited state $E_1$ are small ($\Delta(E_1 - E_0)/(E_1 - E_0)_{\min} \approx 0.23$). Figure 1(d) shows that as the domain wall center for the disordered system is displaced, spatial localization of the qubit is preserved.

General features of flying spin qubits emerge from an analysis of the simplest nontrivial case $n_c = 3$ for which exact solutions exist. The most general $n_c = 3$ Hamiltonian in Eq. (1) can be re-parameterized as

$$H_3 = \sum_{k=1}^{3} \left( \tilde{J}_0 + \tilde{J}_1 \cos\left(2\pi(k-1)/3 - \varphi\right) \right) \mathbf{S}_k \cdot \mathbf{S}_{k+1}, \qquad (2)$$

where $\tilde{J}_0$ and $\tilde{J}_1$ are constants, and $\varphi$ represents the phase of the domain wall around the ring. The $(S, S_Z) = (1/2, 1/2)$ subspace is two-dimensional, spanned by the states $|\pm\rangle \equiv |001\rangle + e^{\pm 2\pi i/3}|010\rangle + e^{\pm 4\pi i/3}|100\rangle$. Using this basis, we can re-express $H_3(\varphi) = \Delta(\Sigma_X \cos\varphi + \Sigma_Y \sin\varphi)$, where the energy gap between the ground and first excited states is $\Delta = 3\tilde{J}_1/4$ and $\Sigma_{X \pm iY} \equiv \frac{4}{3} \sum_{k=1}^{3} e^{\pm 2\pi i(k-1)/3} \mathbf{S}_k \cdot \mathbf{S}_{k+1}$. Together with $\Sigma_Z = -\frac{i}{2}[\Sigma_X, \Sigma_Y]$, the operators $\{\Sigma_X, \Sigma_Y, \Sigma_Z\}$ satisfy $[\Sigma_a, \Sigma_b] = 2i\varepsilon_{abc}\Sigma_c$, and the states $\{|\pm\rangle\}$ can be regarded as a pseudo-spin doublet (Figure 2a). Spatially uniform exchange (parameterized by $\tilde{J}_0$) does not couple to these states, and the ground state energy of $H_3(\varphi)$ *is independent of* $\varphi$. We can interpret $\{|\pm\rangle\}$ as right- and left-traveling spin-current Bloch states, eigenstates of the discrete translation operator $D_2$ (defined by $D_2 \mathbf{S}_k \equiv \mathbf{S}_{k+2} D_2$) over two lattice sites. The ground state of $H_3(\varphi)$ is given by $|\varphi\rangle \equiv |+\rangle - e^{i\varphi}|-\rangle$. The case $\varphi = 0$ yields explicitly $|\varphi=0\rangle = |010\rangle - |100\rangle \equiv |SS0\rangle$, which corresponds to spins 1 and 2 being in a singlet state, and spin 3 being in the "0" state. Adiabatic evolution of $\varphi$ coherently moves the spin qubit around the ring, such that $|\varphi = 2\pi/3\rangle = |0SS\rangle$ and $|\varphi = 4\pi/3\rangle = |S0S\rangle$. Note that these three states are *not* mutually orthogonal—they cannot be, since the space in which they evolve is two-dimensional.

By letting $\varphi = \omega t$, the domain wall can be moved at a constant speed (Figure 2b). Because the Hamiltonian is now explicitly time-dependent, there are no longer stationary states; however, one may employ Bloch-Floquet theory[19] to understand the steady-state dynamics. The time evolution is governed by a unitary operator $U_t$ that satisfies the Schrödinger equation $i\partial_t U_t = H(t)U_t$, subject to the initial condition $U_0 = \mathbf{1}$. Floquet states are defined here to be the eigenstates of the combined time and space translation operator $F = D_2 U_{4\pi/n_c\omega}$. Full translation around a closed spin chain by two revolutions (governed by the operator $F^{n_c}$ or equivalently $U_{4\pi/\omega}$) yields the same Floquet states as for $F$.

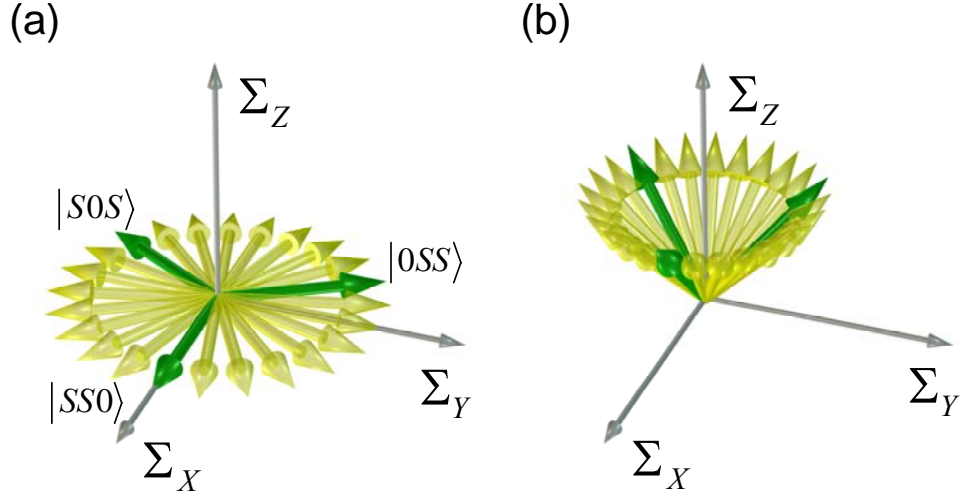

**Figure 2: Analytical model.** (a) Coherent evolution in the pseudo spin-1/2 space for a $n_c = 3$ spin ring. Adiabatic evolution transports the spin state coherently between three non-orthogonal states shown as green arrows. (b) At high domain wall velocities the Floquet states become distorted and eventually merge with eigenstates of $\Sigma_Z$.

For the three-spin ring, we obtain the remarkable *exact* result for the Floquet state associated with the ground state:

$$|X+;\pm\omega\rangle = |SS0\rangle + \frac{\omega}{\Delta}|\pm\rangle, \omega > 0. \tag{3}$$

This state can be interpreted as a "snapshot" of the steady-state quantum dynamics at intervals in time $t = 0, 2\pi/\omega, 4\pi/\omega,...$, and is valid for all $\omega$. One can compute exactly the geometric (Berry) phase associated with these dynamics $\Omega = 2\pi\left(1 - 1/\sqrt{1+(\Delta/\omega)^2}\right)$.

The adiabatic regime can be defined to be the range $|\omega| < \Delta$, in accordance with the energy-time uncertainty principle.

The exact results obtained for the three-spin ring extrapolate well to larger systems. To demonstrate, we consider a $n_c = 9$ spin ring with a single domain wall given by $J_k = J_0 + (-1)^k \alpha(k - k_0)$, where $J_0 = 1$ and $\alpha(k - k_0) = a_0 \sin(\pi(k - k_0)/n_c - \omega t)$. With $k_0 \to k_0 + \frac{n_c}{\pi} \omega t$, this staggered order parameter $\alpha$ corresponds to

$$\alpha(k - k_0) = a_0 \sum_{r=-\infty}^{\infty} (-1)^r \tanh((k - (k_0 + r\, n_c))/w)$$ in the limit $w \to n_c$.

Choosing $a_0 = 0.1$ and $k_0 = n_c/2$, we obtain the Floquet states of the operator $D_2 U_{2\pi/n_c\omega}$ for the $n_c = 9$ spin ring. The ground state $|\varphi_s^0\rangle$ is the analogue of the state $|SS0\rangle$ in Eq. (3), where each spin in Eq. (3) has been replaced by a three-spin cluster qubit[20]. The state $|\varphi_s^0\rangle$ is determined by numerical diagonalization of the initial Hamiltonian. We also find that the Floquet state $|\varphi_s(\omega)\rangle$ for finite ω is well-approximated (to within 1%) by the following expression

$$|\varphi_s^{th}(\omega)\rangle = |\varphi_s^0\rangle + \frac{\omega}{\Delta E}|+\rangle, \qquad (4)$$

similar to the form given in Eq. (3). Here, $|+\rangle = \frac{1}{\sqrt{2}}(|u_1^0\rangle + i|u_2^0\rangle)$ is an eigenstate of $D_2$ for the $n_c = 9$ spin ring, and $|u_1^0\rangle$ and $|u_2^0\rangle$ are the ground states of the Hamiltonian in Eq. (1) with $n_c = 9$ and $J_k = 1$ for all $k$. The energy gap $\Delta E$ between the ground state and the first excited state is finite and is independent of the position of the domain wall.

We now discuss a method of generating EPR pairs from a fully dimerized spin chain. Numerical investigations are performed for a closed chain with $n_c = 30$ spins, using Lanczos diagonalization to determine the spin density and energies. The system is assumed to be initialized in the spin-singlet ground state ($S = 0$). Two domain walls, denoted $A$ and $B$, are created and moved in opposite directions (Figure 3a). The exchange profile describing this system is given by

$J_k = 0.55 - 0.45(-1)^k \left[1 + \alpha(k - k_A) - \alpha(k - k_B)\right]$, where $k_A = (n_c + s)/2$,

$k_B = (n_c - s)/2$, and $s$ is the separation between the two domain walls. Note that the exact quantitative form of the profile is not crucial to the method of producing EPR pairs, provided the general characteristics which produce the domain walls are retained in the exchange profile.

Because the exchange interaction conserves $S$, the two qubits associated with the domain walls must exist in a spin singlet or EPR pair: $|\psi_0\rangle = \frac{1}{\sqrt{2}} \left( |\uparrow\rangle_A |\downarrow\rangle_B - |\downarrow\rangle_A |\uparrow\rangle_B \right)$. In order to visualize this state, one can hybridize it with the first excited (triplet) state

$|\psi_1\rangle = \frac{1}{\sqrt{2}} \left( |\uparrow\rangle_A |\downarrow\rangle_B + |\downarrow\rangle_A |\uparrow\rangle_B \right)$ and compute the spin density for

$(|\psi_0\rangle - |\psi_1\rangle)/\sqrt{2} = |\downarrow\rangle_A |\uparrow\rangle_B$ (Figure 3b). The energies of the three lowest states behave as expected: there is an exchange splitting $\Delta E = E_1 - E_0$ for the two spin qubits which decreases exponentially as the domain walls are moved apart (Figure 3c). The next excited state (energy $E_2$) is given approximately by the one-magnon gap energy[21], which is largely unaffected by the domain wall states (i.e., $E_2 - E_0$ is approximately constant). Within a larger system it would be possible to "radiate" multiple EPR pairs. Because they are entangled states, EPR pairs constitute an important physical resource for applications such as quantum teleportation[13].

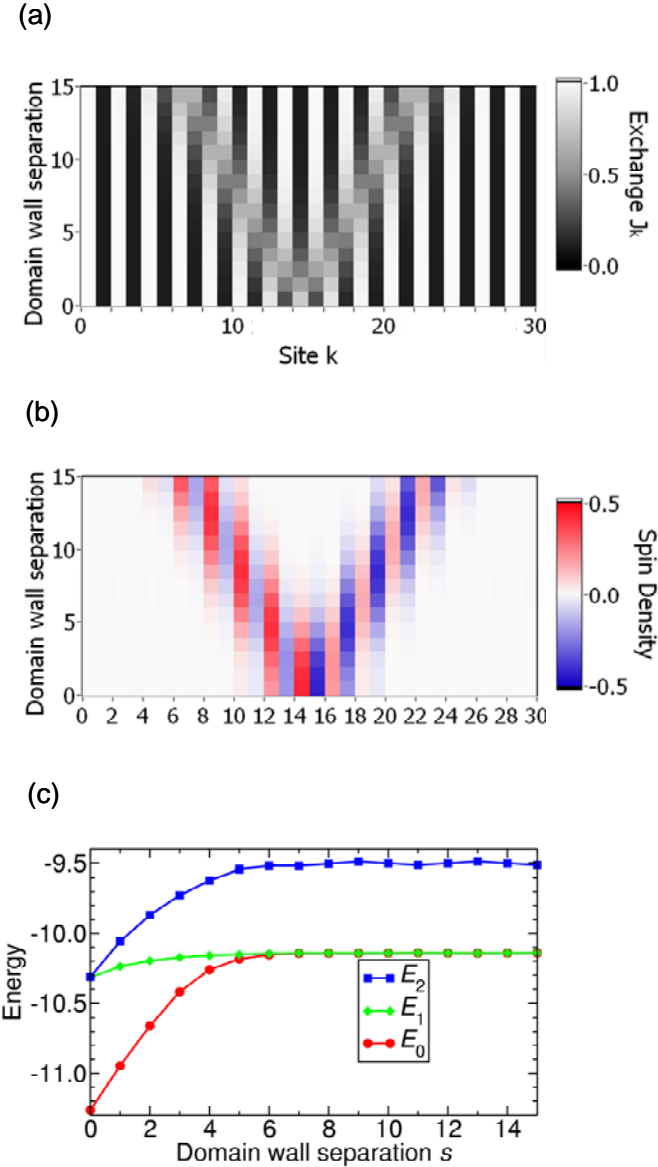

**Figure 3: Mechanism for EPR pair generation.** (a) Exchange profile for a $n_c = 30$ spin chain that is initially dimerized uniformly ($J_{k=odd} = 1.0$, $J_{k=even} = 0.1$). A domain wall pair is produced, and the walls move outward, creating an entangled pair of soliton-like states. (b) Spin density for a linear combination of the (spin-up) ground and first excited states, showing spatial separation of the qubit states as expected. (c) Lowest three energies (in units of $J$) as a function of the domain wall separation $s$, showing the expected exchange splitting between $E_0$ and $E_1$, and a relatively constant spin-wave gap between $E_0$ and $E_2$.

The proposed mechanism for flying spin qubits must be capable of implementation in order to be relevant for quantum computing architectures. Here we describe a specific realization using a one-dimensional array of elliptically-shaped quantum dots, each containing one spin-1/2 electron (Figure 4). Application of an electric field transverse to the array axis modulates the exchange interaction strength between each pair of nearest-neighbor quantum dots (Figure 4a). Each zero-crossing of the electric field corresponds to a single (flying) spin qubit (Figure 4b). The "pseudo-digital" nature[22] of $J(E)$, dependent on the detailed shape of the quantum dots, produces qubits that become more localized with increasing electric field amplitude. The electric field required to transport flying spin qubits may be implemented in a variety of ways, e.g. using a suitably designed coplanar waveguide.

The channel capacity $Q$ of such a device can be estimated in terms of the nominal exchange strength $J$ between nearest-neighbor dots and the domain wall width $w$. The energy gap for the qubit states scales as $\Delta \sim J/w$, similar to that for spin cluster qubits[20]. The qubit can travel $w$ sites in a time $\hbar/\Delta$ without violating the energy-time uncertainty principle. Taking $D \approx 100$ sites for the spacing between domain walls, and using parameters relevant to Ge/Si quantum dots[23] separated by $d$=35 nm and coupled by direct exchange $J_0 \sim 500\ \mu\text{eV}$ [24], one obtains $Q = J_0/\hbar D\ qubit \approx 7.6 \times 10^9\ qubit/\text{s}$. Limitations on the switching speed for the electric field may reduce the actual channel capacity from this maximum value; however, the mechanism of electric field propagation itself does not limit the channel capacity.

Flying spin qubits should prove useful at all architectural levels, such as transporting "fresh qubits" for quantum error correction, carrying qubits to readout locations, implementing quantum gating between remote qubits, and other tasks.

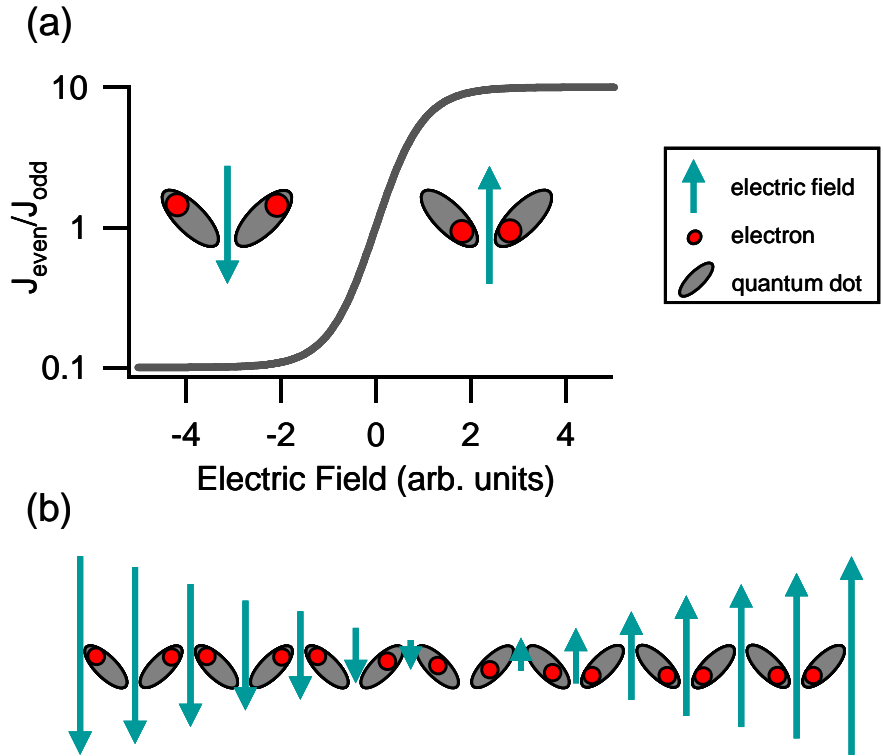

**Figure 4: Experimental realization for flying spin qubits.** (a) Dimerized exchange $J(E)$ produced by elliptically-shaped quantum dots in a transverse electric field. The field profile maps directly onto the staggered order parameter $\alpha$. (b) Zero-crossings of the electric field correspond to single qubit states.

In summary, we have demonstrated a mechanism by which flying spin qubits can be produced and moved controllably entirely within the solid state. Rapid, high fidelity transport of spin qubits is achieved by designing a quantum field with soliton-like states that support localized spin states. The qubits created in this manner are topologically stable against a wide range of disturbances. By creating pairs of domain walls from a uniformly dimerized state, EPR pairs may be efficiently generated. Finally, we have proposed a scheme for implementing flying spin qubits in an array of elliptically-shaped quantum dots.

We thank Yaakov Weinstein for helpful comments. Computations were performed at the ASC DoD Major Shared Resource Center.

## References

1. D. P. DiVincenzo, Fortschr. Phys. **48**, 771 (2000).
2. D. Loss and D. P. DiVincenzo, Phys. Rev. A **57**, 120 (1998).
3. B. E. Kane, Nature **393**, 133 (1998).
4. V. Privman, I. D. Vagner, and G. Kventsel, Phys. Lett. A **239**, 141 (1998).
5. A. Imamoglu, D. D. Awschalom, G. Burkard, et al., Phys. Rev. Lett. **83**, 4204 (1999).
6. R. Vrijen, E. Yablonovitch, K. Wang, et al., Phys. Rev. A **62**, 012306 (2000).
7. J. Levy, Phys. Rev. A **64**, 052306 (2001).
8. T. D. Ladd, J. R. Goldman, F. Yamaguchi, et al., Phys. Rev. Lett. **89**, 017901 (2002).
9. S. Bose, Phys. Rev. Lett. **91**, 207901 (2003).
10. M. Christandl, N. Datta, A. Ekert, et al., Phys. Rev. Lett. **92**, 187902 (2004).
11. A. Wojcik, T. Luczak, P. Kurzynski, et al., Phys. Rev. A **72**, 034303 (2005).
12. G. De Chiara, D. Rossini, S. Montangero, et al., Phys. Rev. A **72**, 012323 (2005).
13. C. H. Bennett, G. Brassard, C. Crepeau, et al., Phys. Rev. Lett. **70**, 1895 (1993).
14. A. J. Heeger, S. Kivelson, J. R. Schreiffer, et al., Rev. Mod. Phys. **60**, 781 (1988).
15. J. Cullum and R. A. Willoughby, *Lanczos Algorithms for Large Symmetric Eigenvalue Computations* (Birkhauser, Boston, 1985).
16. A. Auerbach, *Interacting electrons and quantum magnetism* (Springer-Verlag, New York, 1994).
17. P. Jordan and E. P. Wigner, Z. Phys. **47**, 631 (1928).
18. T. F. Nakano, H, J. Phys. Soc. Japan **49**, 1679 (1980).
19. J. H. Shirley, Phys. Rev. **138**, B979 (1965).
20. F. Meier, J. Levy, and D. Loss, Phys. Rev. Lett. **90**, 047901 (2003).
21. T. Barnes, J. Riera, and D. A. Tennant, Phys. Rev. B **59**, 11384 (1999).
22. M. Friesen, R. Joynt, and M. A. Eriksson, Appl. Phys. Lett. **81**, 4619 (2002).
23. O. Guise, J. Ahner, J. John T. Yates, et al., Appl. Phys. Lett. **87**, 171902 (2005).
24. Unpublished result from C. Pryor, et al.